\magnification=\magstep1


\def\sd{standard deviation}

\def\pd{probability distribution}

\def\mapexpl{\ (If $\dot x =0$ at $x=x_1$ then the next value of $x$ for
which $\dot x =0$ is $x_2$.)\ }
\def\damtp{\baselineskip=10pt
\eightsl Department of Applied Mathematics and Theoretical Physics,\break%
\eightsl University of Cambridge, Cambridge CB3 9EW, UK\par}
\def\ont{\baselineskip=10pt
\eightsl Optique nonlin\'eaire th\'eorique, \break
Universit\'e libre de Bruxelles, CP231 \break
Bruxelles 1050 BELGIUM}
\def\and{\ifnyasrefstyle \&\ \else and \fi}%
\def\regime{r\'egime}%
%
%
\def\xm{x_{\rm max}}

\def\xhc{\hat x_{\rm c}}

\def\mle{\mu |\ln \epsilon |}
\def\Or#1{{\cal O}(#1)}
\def\mean#1{{\big<#1\big>}}



\font\sixrm=cmr6           \font\sevenrm=cmr7
\font\sixi=cmmi6           \font\seveni=cmmi7
\font\sixsy=cmsy6          \font\sevensy=cmsy7

\font\eightrm=cmr8         \font\tenrm=cmr10
\font\eighti=cmmi8         
\font\eightsy=cmsy8        
\font\eightit=cmti8        \font\tenit=cmti10
\font\eightsl=cmsl8        \font\tensl=cmsl10
\font\eightbf=cmbx8        \font\tenbf=cmbx10
\font\eighttt=cmtt8        

\font\twelverm=cmr12

\font\twelveit=cmti12
\font\twelvesl=cmsl12
\font\twelvebf=cmbx12

\font\fourteenrm=cmr12 at 14pt

\font\fourteenit=cmti12 at 14pt
\font\fourteensl=cmsl12 at 14pt
\font\fourteenbf=cmbx12 at 14pt




\def\pmb#1{\setbox0=\hbox{#1}%
 \kern-.025em\copy0\kern-\wd0%
 \kern.05em\copy0\kern-\wd0%
 \kern-.025em\raise.0433em\box0}
\def\Bigtype{%
\let\rm=\fourteenrm \let\bf=\fourteenbf%
\let\it=\fourteenit \let\sl=\fourteensl%
\baselineskip=1.44\baselineskip \rm}%
\def\bigtype{%
\let\rm=\twelverm \let\bf=\twelvebf%
\let\it=\twelveit \let\sl=\twelvesl%
\baselineskip=1.2\baselineskip \rm}%
\def\medtype{%
\let\rm=\tenrm \let\bf=\tenbf%
\let\it=\tenit \let\sl=\tensl%
\baselineskip=12pt \rm}%
\def\smalltype{%
\let\rm=\eightrm \let\bf=\eightbf%
\let\it=\eightit \let\sl=\eightsl%
\let\tt=\eighttt
\def\cal{\fam2 }%
 \textfont0=\eightrm   \scriptfont0=\sevenrm \scriptscriptfont0=\sixrm
 \textfont1=\eighti    \scriptfont1=\seveni  \scriptscriptfont1=\sixi
 \textfont2=\eightsy   \scriptfont2=\sevensy \scriptscriptfont2=\sixsy
 \textfont3=\tenex     \scriptfont3=\tenex   \scriptscriptfont3=\tenex
 \textfont\itfam=\eightit
 \textfont\slfam=\eightsl
 \textfont\bffam=\eightbf
 \textfont\ttfam=\eighttt
\baselineskip=10pt
\parskip=1pt plus1pt minus 1pt%
\rm}%
\voffset=0truein%
\vsize=9.3truein%
\hoffset=0.1truein%
\hsize=6.0truein%
\parindent=0.375truein%
\widowpenalty=300

\def\spaceandathird{%
\baselineskip=14pt
\parskip=2pt plus2pt minus2pt%
\everydisplay={\openup 1\jot}}%

\def\singlespace{%
\baselineskip=12pt
\parskip=1pt plus1pt minus 1pt%
\everydisplay={\openup 0\jot}}%

\def\closeup{\singlespace}%

\newdimen\pagewidth
\newdimen\pageheight    
\pagewidth=6.0truein
\vsize=9.3truein%
\pageheight=9.3truein
\newif\ifwindowcover%
\windowcoverfalse%
\outer\def\windowcover{\windowcovertrue}%
\newif\ifcam%
\camfalse%
\outer\def\damtpaddress{\camtrue}%
\newif\ifnofigs%
\nofigsfalse%
\newif\ifchapstart
\chapstartfalse
\def\chapstartpage{\global\chapstarttrue}
\newif\ifinappendices%
\inappendicesfalse%
\newif\iffiguresatend%
\newif\ifnumbereqnsseq%
\def\numbereqnsseq{ \numbereqnsseqtrue}%
\newif\ifnumberfigsseq%
\def\numberfigsseq{ \numberfigsseqtrue}%
\newif\ifnumbertablesseq%
\def\numbertablesseq{ \numbertablesseqtrue}%
\newif\ifnumberrefsintext%
\def\numberrefsintext{ \numberrefsintexttrue}%
\newif\ifinrefs%
\inrefsfalse%
\newif\ifpaperstyle%
\outer\def\paperstyle{\paperstyletrue
     \refauthorinitialssurname%
     \numbereqnsseq \numberfigsseq \numbertablesseq%
     \numberrefsintext}%
\newif\ifsendstyle%
\newif\ifnyasrefstyle%

\newif\ifsinglesided%
\newif\ifdated%

\newif\ifsastyle%
\outer\def\sastyle{\sastyletrue\numberrefsintext%
\ifsinglesided\headline={\rightheadline}
\else\headline={\ifodd\pageno\rightheadline
                        \else\leftheadline\fi}\fi%
\footline={\ifdated\hfil\eightrm\today\else{}\fi}}
\newif\ifheadset%
\headsetfalse%

\countdef\sectno=1%
\countdef\subsectno=2%
\countdef\subsubsectno=3%
\newcount \equano%
\newcount \figno%
\newcount \tableno%
\newcount \referencecount%
\newbox\TitlePage%
\newbox\DedicationPage%
\newbox\PrefacePage%
\newbox\ConclusionPage%
\newbox\AcknowledgementsPage%
\newbox\SummaryPage%
\newbox\TableOfContents%
\newbox\ListOfFigures%
\newbox\Nomenclature%
\newbox\ListOfTables%
\newbox\FiguresPage%
\newbox\figurebox%
\def\thesectno{%
\ifinappendices%
  \ifcase\sectno\or%
    A\or B\or C\or D\or E\or F\or G\or H\or I\or J\or K\or L\or M%
    N\or O\or P\or Q\or R\or S\or T\or U\or V\or W\or X\or Y\or Z\fi%
\else%
  \the\sectno%
\fi}%
\def \lastsubsectno{\thesectno .\the\subsectno}%
\def \lastsubsubsectno{\thesectno .\the\subsectno .\the\subsubsectno}%
\def \lasteqno{\ifnumbereqnsseq (\the\equano)%
\else (\thesectno.\the\equano)\fi}%
\def \preveqno{\advance \equano by -1 \lasteqno}%
\def \lastfigno{\ifnumberfigsseq \hbox{Figure}~\hbox{\the\figno}%
\else \hbox{Figure}~\hbox{\thesectno -\the\figno}\fi}
\def \lasttableno{\ifnumbertablesseq \hbox{Table}~\hbox{\the\tableno}%
\else \hbox{Table}~\hbox{\thesectno .\the\tableno}\fi}%
\def \nextsubsectno{{\advance \subsectno by 1 \lastsubsectno}}
\def \nextsubsubsectno{{\advance \subsubsectno by 1 \lastsubsubsectno}}%
\def \nexteqno{{\advance \equano by 1 \lasteqno}}%
\def \nextfigno{{\advance \figno by 1 \lastfigno}}%
\def \nextnextfigno{{\advance \figno by 2 \lastfigno}}%
\def \nexttableno{{\advance \tableno by 1 \lasttableno}}%
\def\minimalsectno{\ifnum\subsubsectno>0 \lastsubsubsectno%
\else\ifnum\subsectno>0 \lastsubsectno%
\else \thesectno \fi\fi}%
\def\accretetoTOCone#1#2{\begingroup\closeup\raggedright\nohyphens%
  \global\setbox\TableOfContents=\vbox{%
    \unvbox\TableOfContents%
  \filbreak\medskip%
    \hangindent\parindent%
    \textindent{\bf#1}{\bf\strut #2\strut}\hfill\bf\folio\par%
    \smallskip%
  }\endgroup}%

\def\accretetoTOCtwo#1#2{\begingroup\closeup\closeup\raggedright\nohyphens%
  \global\setbox\TableOfContents=\vbox{%
    \unvbox\TableOfContents%
\goodbreak
    \par\indent\hangindent2\parindent%
    \textindent{\sl #1}{\sl\strut #2\strut}\leaderfill\eightrm\folio\par%
  }\endgroup}%

\def\accretetoTOCthree#1#2{\begingroup\closeup\closeup\closeup\raggedright\nohyphens%
  \global\setbox\TableOfContents=\vbox{%
    \unvbox\TableOfContents%
    \par\indent\hbox to\parindent{\hfil}\hangindent3\parindent%
    \textindent{\eightrm #1}{\eightrm\strut #2\strut}\par%
  }\endgroup}%
\def\beginsect#1\par{\begingroup
\ifsastyle\thesisheading{#1}
        \chapstartpage
    \accretetoTOCone{\ifinrefs\else\thesectno.\fi}{#1}%
    \xdef\rhead{\ #1}
\else%
    \global \advance \sectno by 1%
    \mainheading{#1}%
    \accretetoTOCone{\ifinrefs\else\thesectno.\fi}{#1}%
\fi
\global \subsectno=0 \global \subsubsectno=0%
\ifnumbereqnsseq \else \global \equano=0 \fi%
\ifnumberfigsseq \else \global \figno=0 \fi%
\ifnumbertablesseq \else \global \tableno=0 \fi%
\message{\thesectno . #1}%
\nobreak\endgroup%
\ifsastyle\headsetfalse\else\fi}%

\def\saheading#1{\vfill\supereject
        \global \advance \sectno by 1%
        \chapstartpage
        \vskip 0pt plus.3\vsize \penalty-100 \vskip 0pt plus-.3\vsize%
        \ifinrefs\noindent{\bigtype \bf #1}
        \else\ifinappendices\noindent{\bigtype \bf #1}
        \else\noindent{\bigtype \the\sectno .\ \bf #1}\fi\fi
        \par\nobreak\noindent}

\def\thesisheading#1{\vfill\supereject
        \global \advance \sectno by 1%
        \vskip 0pt plus.3\vsize \penalty-100 \vskip 0pt plus-.3\vsize%
        \ifinrefs\noindent{\bigtype \bf #1}
        \else\ifinappendices\noindent{\bigtype \bf Appendix \thesectno .\ #1}
        \else\line{}\vskip 0.5truein
        \centerline{\twelverm CHAPTER \the\sectno}
                \vskip 0.5truein
        \centerline{\fourteenbf #1}
        \vskip 0.5truein
        \fi\fi
        \par\nobreak\noindent}

\def\mainheading#1{\begingroup\nohyphens\hbadness=3000%
        \ifinrefs%
          \noindent {\bf #1}\nobreak%
        \else%
           \ifinappendices%
              \noindent {\bf #1}\par\nobreak%
           \else
              \noindent {\bf \thesectno . #1}\par\nobreak%
           \fi%
        \fi%
    \noindent\endgroup}%

\def\leftheadline{\ifheadset%
\hbox to \pagewidth{%
        \vbox to 10pt{ }
        \llap{\tenrm\folio\kern2pc}
        \kern-1pc\eightit\lhead\hrulefill}\else\line{}\fi}
\def\rightheadline{\ifheadset%
\hbox to \pagewidth{%
        \vbox to 10pt{ }
        \hrulefill\eightrm\thesectno.\eightit\rhead
        \rlap{\kern1pc\tenrm\folio}}\else\line{}\fi}

\def\subheading#1{\begingroup\nohyphens\hbadness=3000%
\bigskip
    \noindent {\sl\bf\lastsubsectno . #1}\par\nobreak\noindent%
\endgroup}%

\def\subsubheading#1{\begingroup\nohyphens\hbadness=3000%
  \vskip 0pt plus.1\vsize \penalty-100 \vskip 0pt plus-.1\vsize%
\medskip\vskip\parskip%
  \noindent {\sl\lastsubsubsectno . #1}\par\nobreak\smallskip\noindent%
\endgroup}%

\outer \def \beginappendix#1\par{%
\inrefsfalse%
\ifinappendices\else\inappendicestrue\global\sectno=0\fi%
\begingroup%
\ifsastyle\saheading{#1}%
\xdef\rhead{#1}
\else%
    \global \advance \sectno by 1%
    \mainheading{#1}%
    \ifnumbereqnsseq \else \global \equano=0 \fi%
    \ifnumberfigsseq \else \global \figno=0 \fi%
    \ifnumbertablesseq \else \global \tableno=0 \fi%
    \global \subsectno=0 \global \subsubsectno=0%
\fi%
\message{\thesectno . #1}%
\accretetoTOCone{}{#1}%
\endgroup%
\ifsastyle\headsetfalse\else\fi%
}

\outer \def \beginsubsect#1\par{
\global \advance \subsectno by 1%
\global \subsubsectno=0%
\accretetoTOCtwo{\lastsubsectno.}{#1}%
\subheading{#1}}

\outer \def \beginsubsubsect#1\par{\global \advance \subsubsectno by 1%
\accretetoTOCthree{\lastsubsubsectno.}{#1}\subsubheading{#1}}%
\def \myeqno{\global \advance \equano by 1 {\rm \lasteqno}}%
\def \myeqlabel#1{\expandafter\xdef\csname #1\endcsname{\lasteqno}}

\def \mytextlabel#1{\expandafter\xdef\csname #1\endcsname{\minimalsectno}}
\def \myChapterlabel#1#2{%
 \expandafter\xdef\csname #1\endcsname{Chapter~#2}}
\def \myAppendixlabel#1#2{%
 \expandafter\xdef\csname #1\endcsname{Appendix~\hbox{#2}}}
\def \myfigurelabel#1{\expandafter\xdef\csname #1\endcsname{\lastfigno}}
\def \mytablelabel#1{\expandafter\xdef\csname #1\endcsname{\lasttableno}}
\input epsf

\def\figure #1 #2 #3\par{\global \advance \figno by 1%
\iffiguresatend%
   \global\setbox\ListOfFigures=\vbox{\unvbox\ListOfFigures%
   \bigskip\caption{\hbox{\bf\lastfigno. }{#3}}
   \par\filbreak}%
   \ifnofigs%
   \else
      \global\setbox\FiguresPage=\vbox{\unvbox\FiguresPage%
        \epsfysize=#1
        \centerline{\epsffile{#2}}
       \vskip 0.5truein\relax%
       \hbox{\bf \lastfigno}
        {#3}
        \vfill\eject}
   \fi%
\else%
\ifsastyle
   \global\setbox\ListOfFigures=\vbox{\unvbox\ListOfFigures%
   \bigskip\caption{\hbox%
{\bf\lastfigno. }
{page \folio},
{size in text #1},
{file #2}\filbreak\noindent{#3}}
   \par\filbreak}%
\fi
  \setbox\figurebox=\vbox{%
  \ifnofigs
  \else
        \epsfysize=#1   
        \centerline{\epsffile{#2}}
  \fi
    \caption{{\bf\lastfigno. }{\eightsl #3}}}
  \topinsert\unvbox\figurebox\endinsert%
\fi}%

\def\oldfigure #1 #2 #3\par{\global \advance \figno by 1%
\iffiguresatend 
   \begingroup%
   \global\setbox\ListOfFigures=\vbox{\unvbox\ListOfFigures%
   \bigskip\caption{\hbox{\bf\lastfigno. }{#3}}
   \par\filbreak}%
   \ifnofigs
   \else
      \global\setbox\FiguresPage=\vbox{\unvbox\FiguresPage%
        \epsfysize=#1
        \centerline{\epsffile{#2}}
       \vskip 0.5truein\relax%
       \hbox{\bf \lastfigno}
        \vfill\eject}
   \fi
   \endgroup%
\else%
  \setbox\figurebox=\vbox{%
  \ifnofigs
  \else
        \epsfysize=#1   
        \centerline{\epsffile{#2}}
  \fi
    \caption{{\bf\lastfigno. }{\eightsl #3}}}
  \topinsert\unvbox\figurebox\endinsert%
\fi}%

\def\twofigures #1 #2 #3 #4\par{\global \advance \figno by 1%
\iffiguresatend 
   \begingroup%
   \global\setbox\ListOfFigures=\vbox{\unvbox\ListOfFigures%
   \bigskip\caption{\hbox{\bf\lastfigno. }{#4}}
   \par\filbreak}%
   \ifnofigs
   \else
      \global\setbox\FiguresPage=\vbox{\unvbox\FiguresPage%
        \epsfysize=#1
        \centerline{\epsffile{#2}\hfil\epsffile{#3}}
       \vskip 0.5truein\relax%
       \hbox{\bf \lastfigno}
        \vfill\eject}
   \fi
   \endgroup%
\else%
  \setbox\figurebox=\vbox{%
  \ifnofigs
  \else
        \centerline{
        \epsfysize=#1   
        (a)\epsffile{#2}
        \ \ 
        \epsfysize=#1   
        (b)\epsffile{#3}}
  \fi
    \caption{{\bf\lastfigno. }{\eightsl #4}}}
  \topinsert\unvbox\figurebox\endinsert%
\fi}%

\def\fourfigures #1 #2 #3 #4 #5 #6\par{\global \advance \figno by 1%
\iffiguresatend 
   \begingroup%
   \global\setbox\ListOfFigures=\vbox{\unvbox\ListOfFigures%
   \bigskip\caption{\hbox{\bf\lastfigno. }{#6}}
   \par\filbreak}%
   \ifnofigs
   \else
      \global\setbox\FiguresPage=\vbox{\unvbox\FiguresPage%
        \centerline{
        \epsfysize=#1   
        (a)\epsffile{#2}
        \ \ 
        \epsfysize=#1   
        (b)\epsffile{#3}}
        \line{ }
        \centerline{
        \epsfysize=#1   
        (c)\epsffile{#4}
        \ \ 
        \epsfysize=#1   
        (d)\epsffile{#5}}
        \hbox{\bf \lastfigno}
         {#6}
        \vfill\eject}
   \fi
   \endgroup%
\else%
  \setbox\figurebox=\vbox{%
  \ifnofigs
  \else
        \centerline{
        \epsfysize=#1   
        (a)\epsffile{#2}
        \ \ 
        \epsfysize=#1   
        (b)\epsffile{#3}}
        \line{ }
        \centerline{
        \epsfysize=#1   
        (c)\epsffile{#4}
        \ \ 
        \epsfysize=#1   
        (d)\epsffile{#5}}
  \fi
    \caption{{\bf\lastfigno. }{\eightsl #6}}}
  \topinsert\unvbox\figurebox\endinsert%
\fi}%

\def\eightfigures #1 #2 #3 #4 #5 #6 #7 #8 #9%
\par{\global \advance \figno by 1%
\iffiguresatend 
   \begingroup%
   \global\setbox\ListOfFigures=\vbox{\unvbox\ListOfFigures%
   \bigskip\caption{\hbox{\bf\lastfigno. }{#6}}
   \par\filbreak}%
   \ifnofigs
   \else
      \global\setbox\FiguresPage=\vbox{\unvbox\FiguresPage%
        \epsfysize=#1
        \centerline{\epsffile{#2}\hfil\epsffile{#3}}
       \vskip 0.5truein\relax%
       \hbox{\bf \lastfigno}
        \vfill\eject}
   \fi
   \endgroup%
\else%
  \setbox\figurebox=\vbox{%
  \ifnofigs
  \else
        \centerline{
        \epsfysize=1.8truein
        (a)\epsffile{#1}
        \epsfysize=1.8truein
        (b)\epsffile{#2}
        \epsfysize=1.8truein
        (c)\epsffile{#3}}
        \line{ }
        \centerline{
        \epsfysize=1.8truein
        (d)\epsffile{#4}
        \epsfysize=1.8truein
        (e)\epsffile{#5}
        \epsfysize=1.8truein
        (f)\epsffile{#6}}
        \line{ }
        \centerline{
        \epsfysize=1.8truein
        (g)\epsffile{#7}
        \epsfysize=1.8truein
        (h)\epsffile{#8}
        \epsfysize=1.8truein
        \ \ \ \epsffile{key.eps}}
  \fi
    \caption{{\bf\lastfigno. }{\eightsl #9}}
}
  \topinsert\unvbox\figurebox\endinsert%
\fi}%

\def\caption#1{%
\smallskip%
\begingroup
\nohyphens%
\advance \leftskip by \parindent%
\advance \rightskip by \parindent
\spaceskip=.3333em plus .3333em \xspaceskip=.5em plus .5em%
\par \noindent{%
\iffiguresatend{#1}\else\smalltype{#1}\par\fi\par} \par\endgroup}%
\def\nom#1#2{\begingroup\closeup%
\global\setbox\Nomenclature=\vbox{\unvbox\Nomenclature\smallskip%
\parindent=7em\item{#1}\strut\hskip0.2truein #2\strut\par}\endgroup}%
\def\toindex#1#2{\begingroup%
\ifundefined{#1index}%
\expandafter\xdef\csname#1index\endcsname{#2\hfill\folio}%
\else
\expandafter\xdef\csname#1index\endcsname{\csname#1index\endcsname, \folio}%
\fi%
\endgroup}
\def\shoindexentry#1{\ifundefined{#1index}\else%
\smallskip{\tenrm\noindent{\csname#1index\endcsname}}\fi}
\outer \def \beginrefs{\inrefstrue\inappendicesfalse}%

\def\showrefs{\ifsastyle\accretetoTOCone{}{References}\headsetfalse\else\fi%
\setbox\myrefs=\vbox{\unvbox\myrefs\vfil}%
\ifsendstyle\vfill\eject\else\smallskip\goodbreak\fi
    \unvbox\myrefs}

\def\addtorefs#1#2{%
\global\advance\referencecount by 1%
\expandafter\xdef\csname#2\endcsname{\the\referencecount}%
\global\setbox\myrefs=\vbox{\unvbox\myrefs%
{\ifsendstyle\else\closeup\fi
\ifnyasrefstyle
\item{\csname#2\endcsname.}\frenchspacing{\csname#1\endcsname}
\else
\item{[\csname#2\endcsname]}\frenchspacing{\csname#1\endcsname}
\fi
\hfil\par\ifsendstyle\medskip\fi}}}

\def\addref#1#2{\ifundefined{#2}\addtorefs{#1}{#2}\fi
\ifnyasrefstyle$^{\csname#2\endcsname}$\else~[\csname#2\endcsname]\fi}%


\def\addtworefs#1#2#3#4{%
{\ifundefined{#2}\addtorefs{#1}{#2}\fi}%
{\ifundefined{#4}\addtorefs{#3}{#4}\fi}%
\ifnyasrefstyle
$^{\csname#2\endcsname,~\csname#4\endcsname}$%
\else~[\csname#2\endcsname,~\csname#4\endcsname]\fi}%

\def\addthreerefs#1#2#3#4#5#6{%
{\ifundefined{#2}\addtorefs{#1}{#2}\fi}%
{\ifundefined{#4}\addtorefs{#3}{#4}\fi}%
{\ifundefined{#6}\addtorefs{#5}{#6}\fi}%
\ifnyasrefstyle
$^{\csname#2\endcsname,\csname#4\endcsname,\csname#6\endcsname}$%
\else~[\csname#2\endcsname,\csname#4\endcsname,\csname#6\endcsname]\fi}%

\def\addfourrefs#1#2#3#4#5#6#7#8{
{\ifundefined{#2}\addtorefs{#1}{#2}\fi}%
{\ifundefined{#4}\addtorefs{#3}{#4}\fi}%
{\ifundefined{#6}\addtorefs{#5}{#6}\fi}%
{\ifundefined{#8}\addtorefs{#7}{#8}\fi}%
\ifnyasrefstyle
$^{\csname#2\endcsname-\csname#8\endcsname}$%
\else~[\csname#2\endcsname-\csname#8\endcsname]\fi}%


\def\genericref#1#2{\begingroup\closeup%
\expandafter\xdef\csname#1\endcsname{#2.}%
\endgroup}%
\def\refyear#1{\ifnyasrefstyle #1. \else (#1)\fi}%
\def\reftitle#1{\lq\lq#1''}%
\def\refauthorinitialssurname{%
\def\author##1##2{##1~##2}
\def\authorn##1##2{\ifnyasrefstyle\uppercase{##2,~##1}\else{##1~##2}\fi}}%
\refauthorinitialssurname%
\outer\def\refarticle#1#2#3#4#5#6#7{%
\genericref{#1}{%
\ifnyasrefstyle
    {#2.\ \refyear{#3} \reftitle{#4}. {#5} {\bf #6}: {#7}}%
\else%
  \ifpaperstyle%
    {#2\ \reftitle{#4.} {\sl #5} {\bf #6} #7\ \refyear{#3}}%
  \else
    {#2\ \reftitle{#4.} {\sl #5} {\bf #6} #7\ \refyear{#3}}%
  \fi%
\fi}}%

\outer\def\refbook#1#2#3#4#5#6{%
\genericref{#1}%
{\ifnyasrefstyle%
#2.\ #3. #4. #5, #6%
\else%
\ifpaperstyle%
{#2,\ {\sl #4}\ (#5, \ #6, \ #3)}%
  \else%
    {#2\ {\sl #4}.  #5: #6 \ \refyear{#3}}%
\fi\fi}}%

\outer\def\refnote#1{%
\genericref{#1}}
\def\theauthor{Grant Lythe}

\def\title#1\par{\message{#1}%
\xdef\lhead{#1}%
\xdef\rhead{}%
\begingroup
  \ifwindowcover%
    \dimen1=2.5truein \advance\dimen1 by -\voffset%
    \global\setbox\TitlePage=\vbox{\vbox to \dimen1{%
      \dimen2=1.75truein \advance\dimen2 by -\voffset%
      \smallskip \relax%
      \dimen3=1.5truein \advance\dimen3 by -\hoffset%
      \leftskip=\dimen3 plus3em\relax%
      \dimen4=\hsize \advance\dimen4 by \hoffset \advance\dimen4 by -5truein%
      \rightskip=\dimen4 plus3em\relax%
      \parfillskip=0pt \parindent=0pt \spaceskip=0.3333em
        \xspaceskip=0.5em%
        \line{}\bigskip\line{}\bigskip
      \centerline{\bigtype\bf #1}\smallskip
          {\centerline{\theauthor}}
        }}
  \fi%
\endgroup}%

\def\signaturepaper{
\global\setbox\TitlePage=\vbox{%
\unvbox\TitlePage%
\noindent\narrower\raggedcenter{\bigtype%
\ifwindowcover \else \theauthor\par\fi%
\ifcam\damtp\else\ont\fi
\bigskip}}}%

\def\signaturesa{
\global\setbox\TitlePage=\vbox{%
\unvbox\TitlePage%
\noindent\narrower\raggedcenter{\bigtype%
\ifwindowcover \smallskip \else \theauthor\par\fi%
\centerline{\tensl Trinity College,}
\centerline{\tensl Cambridge.}
\centerline{\tenrm July 1994}\vskip 0.5truein}}}%

\def\signaturethesis{%
\parskip=20pt plus 10pt%
\global\setbox\TitlePage=\vbox{%
\line{}\bigskip
\vfill%
\noindent\narrower\raggedcenter{%
\bigtype%
{\bf Stochastic slow-fast dynamics}\par%
\bigtype%
Grant David Lythe\par%
\medtype
Trinity College, Cambridge\par%
\vfill%
Dissertation submitted\break%
for the degree of\break%
Doctor of Philosophy\break%
at the\break%
University of Cambridge\break%
\vfill%
October 1994}}}%

\def\signature{\begingroup%
\ifpaperstyle \signaturepaper \fi%
\ifsastyle \signaturethesis \fi%
\endgroup}%

\long\def\addtotitlepage#1{
\global\setbox\TitlePage=\vbox{%
\unvbox\TitlePage%
\bigskip\noindent#1\par}}%

\def\abstract#1\par{\begingroup%
\addtotitlepage{{\centerline{\bf Abstract}}\par\nobreak\medskip\noindent#1}%
\endgroup}%

\def\submittedto#1{\begingroup\closeup%
\addtotitlepage{\smalltype
\ifpaperstyle  {Submitted to  \hfill printed on \ \today}\fi%
}\endgroup}%

\def\publishedin#1{\begingroup\closeup%
\addtotitlepage{\bigtype
\ifpaperstyle  {#1}\fi}\endgroup}%

\def\dedication#1\par{%
\begingroup%
  \global\setbox\DedicationPage=\vbox{%
  \unvbox\DedicationPage%
  \raggedcenter{\sl#1\par}}%
\endgroup}%
\long\def\preface#1{%
  \global\setbox\PrefacePage=\vbox{%
  \unvbox\PrefacePage \noindent #1\par}%
  \begingroup\pageno=-2\endgroup}%
\long\def\thesisacks#1{%
  \global\setbox\AcknowledgementsPage=\vbox{%
  \unvbox\AcknowledgementsPage \noindent #1\par}%
  \begingroup\pageno=-2\endgroup}%
\long\def\conclusion#1{%
  \global\setbox\ConclusionPage=\vbox{%
  \unvbox\ConclusionPage \noindent #1\par}%
  \begingroup\pageno=-2\endgroup}%

\long\def\summary#1{%
\global\setbox\SummaryPage=\vbox{%
  \unvbox\SummaryPage \noindent #1\par}%
  \begingroup\pageno=-4\endgroup}

\outer \def \beginack{
\noindent{\bf{Acknowledgement}}\break}

\def\Showbox#1{\begingroup%
\loop%
\ifvbox#1%
    \dimen255=\vsize%
    \advance \dimen255 by -\topskip%
    \setbox0=\vsplit#1 to \dimen255%
    \unvbox0%
    \null \vfill \penalty -10000%
\repeat%
\endgroup}%

\def\ShowTP{\begingroup \nopagenumbers\headline={}%
\ifvbox\TitlePage%
    \global \sectno=0 \global \subsectno=0 \global \subsubsectno=0%
    \setbox\TitlePage=\vbox{\unvbox\TitlePage\vfil}%
    \Showbox\TitlePage%
\fi \endgroup}%

\def\PreambleHeader#1{%
\begingroup\raggedcenter\line{}\vskip 0.3truein 
\ifsastyle\bigtype\else\Bigtype\fi\bf \noindent%
#1\par \vskip 0.2truein\endgroup\noindent}%

\def\ShowDed{\begingroup \nopagenumbers%
\ifvbox\DedicationPage%
    \global \sectno=0 \global \subsectno=0 \global \subsubsectno=0%
    \setbox\DedicationPage=\vbox{\line{}\vskip 2truein\relax%
                               \unvbox\DedicationPage \vfil}%
    \Showbox\DedicationPage%
\fi\endgroup}%

\def\ShowPref{%
\ifvbox\PrefacePage%
    \global \sectno=0 \global \subsectno=0 \global \subsubsectno=0%
    \setbox\PrefacePage=\vbox{%
\PreambleHeader{Preface}\unvbox\PrefacePage \vfil}%
    \Showbox\PrefacePage%
\fi}%

\def\ShowConc{%
\ifvbox\PrefacePage%
    \global \sectno=0 \global \subsectno=0 \global \subsubsectno=0%
    \setbox\ConclusionPage=\vbox{%
\unvbox\ConclusionPage \vfil}%
    \Showbox\ConclusionPage%
\fi}%

\def\ShowAck{%
\ifvbox\AcknowledgementsPage%
    \global \sectno=0 \global \subsectno=0 \global \subsubsectno=0%
    \setbox\AcknowledgementsPage=\vbox{%
\PreambleHeader{Acknowledgements}\unvbox\AcknowledgementsPage \vfil}%
    \Showbox\AcknowledgementsPage%
\fi}%

\def\ShowSummary{%
\ifvbox\SummaryPage%
    \global \sectno=0 \global \subsectno=0 \global \subsubsectno=0%
    \setbox\SummaryPage=\vbox{%
\PreambleHeader{Summary}\unvbox\SummaryPage \vfil}%
    \Showbox\SummaryPage%
\fi}%

\def\ShowTOC{%
\ifvbox\TableOfContents%
    \global \sectno=0 \global \subsectno=0 \global \subsubsectno=0%
    \setbox\TableOfContents=\vbox{%
\PreambleHeader{Contents}\unvbox\TableOfContents \vfil}%
    \Showbox\TableOfContents%
\fi}%

\def\ShowLOF{%
\ifvbox\ListOfFigures%
    \setbox\ListOfFigures=\vbox{%
\PreambleHeader{\ifpaperstyle Figure captions \else List of Figures\fi}
\unvbox\ListOfFigures \vfil}%
    \Showbox\ListOfFigures%
\fi}%

\def\ShowFP{%
\ifvbox\FiguresPage%
    \setbox\FiguresPage=\vbox{\unvbox\FiguresPage\vfil}%
    \Showbox\FiguresPage%
\fi}%

\def\ShowLOT{%
\ifvbox\ListOfTables%
    \setbox\ListOfTables=\vbox{%
\PreambleHeader{List of Tables}\unvbox\ListOfTables \vfil}%
    \Showbox\ListOfTables%
\fi}%

%
\def\themonth{\ifcase\month\or%
  January\or February\or March\or April\or%
  May\or June\or July\or August\or%
  September\or October\or November\or December\fi}%
\def\today{\number\day%
\space%
\themonth%
\space%
\number\year}%
%
\outer\def\raggedright{\rightskip=0pt plus 4em%
  \spaceskip=.3333em \xspaceskip=.5em%
  \pretolerance=200 \tolerance=400}%
\def\leaderfill{\leaders\hbox to 1em{\hss.\hss}\hfill}%
\def\dddot#1{\begingroup\vbox{%
\moveleft 0.15em\hbox{...}%
\nointerlineskip\vskip0.4ex%
\hbox{$\displaystyle{#1}$}}\endgroup}%
\def\ifundefined#1{\expandafter\ifx\csname#1\endcsname\relax}%
\hyphenation{co-dimen-sion sol-ution sol-utions sun-spot sun-spots
pro-pen-sity}
\overfullrule=0pt%
\def\nohyphens{\pretolerance=9999 \tolerance=9999%
\hyphenpenalty=9999 \exhyphenpenalty=1000 }%
\def\raggedcenter{\leftskip=0pt plus6em \rightskip=\leftskip%
\parfillskip=0pt \parindent=0pt \spaceskip=0.3333em \xspaceskip=0.5em%
\nohyphens \hbadness=3000}%
%
%
%
%
%
\outer\def \table#1\par{\global \advance \tableno by 1%
\begingroup \closeup
\global\setbox\ListOfTables=\vbox{\unvbox\ListOfTables%
\smallskip\noindent{\bf\lasttableno.\ }{\strut\sl #1\strut}\par}\endgroup%
\medskip\caption{{\bf\lasttableno.\ }{\sl #1}}}%

\paperstyle

\newbox\myrefs%
    \setbox\myrefs=\vbox{%
\ifsastyle\noindent{\bigtype \bf{References}\medskip}
\else\smallskip\noindent\bf References\par\medskip\fi}%

\beginrefs 
\begingroup\raggedright%
\nohyphens%

\refarticle{refH101}
{\author{D.W.}{Hughes} \and \author{M.R.E.}{Proctor}}{1990}
            {Chaos and the effect of noise in a model of three-wave
mode coupling}
            {Physica D}{46}{163--176}

\refarticle{refH102}
{\author{D.W.}{Hughes} \and \author{M.R.E.}{Proctor}}{1990}
            {A low order model of the shear instability of convection:
chaos and the effect of noise}
            {Nonlinearity}{3}{127--153}

\refarticle{refH103}
{\author{M.R.E.}{Proctor} \and\author{D.W.}{Hughes}}{1990}%
{The false Hopf bifurcation and noise sensitivity
        in bifurcations with symmetry}
            {Eur.J.Mech.,B/Fluids}{10}{81--86}

\refarticle{refH104}
{\author{Rebecca L.}{Honeycutt}}{1992}
            {Stochastic Runge-Kutta algorithms. I. White noise}
            {Phys. Rev. A}{45}{600--603}

\refarticle{refB102}
{\author{Jean-Marie}{Wersinger}, \author{John M.}{Finn}
  \and \author{Edward}{Ott}}{1980}
            {Bifurcations and strange behaviour in instability
saturation by nonlinear mode coupling}
         {Phys. Rev. Lett.}{44}{453--457}

\refarticle{refV100}
{\author{S.Ya.}{Vyshkind} \and \author{M.I.}{Rabinovich}}{1976}
            {The phase stochastization mechanism and the structure of
wave turbulence in dissipative media}
            {Sov. Phys. JETP}{44}{292--299}
\refarticle{refW100}
{\author{J.-M.}{Wersinger}, \author{J.M.}{Finn}
  \and \author{E.}{Ott}}{1980}
            {Bifurcation and ``strange'' behaviour in instability
saturation by nonlinear three-wave mode coupling}
            {Phys. Fluids}{23}{1142--1154}

\refnote{refNote1}
{The change of variables from the equations for
the amplitudes of the three wave modes used here differs  from
that of Hughes and Proctor; our $(x,y,z)$ corresponds to $(-x,w,y)$}

\refbook{refK100}
{\author{Peter E.}{Kloeden} \and \author{Eckhard}{Platen}}{1992}
    {Numerical Solution of Stochastic Differential Equations}
    {Springer}{Berlin}

\refarticle{refM100}
{\author{Paul}{Mandel} \and \author{T.}{Erneux}}{1984}
            {Laser Lorenz equations with a time-dependent parameter}
            {Phys. Rev. Lett.}{53}{1818--1820}

\refarticle{refM101}
{\author{Paul}{Mandel} \and \author{Thomas}{Erneux}}{1987}
            {The slow passage through a steady bifurcation:
        delay and memory effects}
            {J. Stat. Phys.}{48}{1059--1070}

\refarticle{refM102}
{\author{R.}{Mannella}, \author{Frank}{Moss}
\and \author{P.V.E.}{McClintock}}{1987}
           {Postponed bifurcations of a ring-laser with a swept parameter
        and additive colored noise}
           {Phys. Rev. A}{35}{2560--2566}
\refbook{refM103}
{\author{Frank}{Moss} \and \author{P.V.E.}{McClintock}}{1987}
            {Noise and nonlinear dynamical systems}
            {Cambridge University Press}{Cambridge}

\refarticle{refM104}
{\author{Miltiades}{Georgiou} \and \author{Thomas}{Erneux}}{1992}
          {Pulsating laser oscillations depend on 
        extremely-small-amplitude noise}
           {Phys. Rev. A}{45}{6636--6642}

\refarticle{refS100}
{\author{Emily}{Stone} \and \author{Philip}{Holmes}}{1990}
            {Random perturbations of heteroclinic attractors}
            {SIAM J. Appl. Math.}{50}{726--743}

\refarticle{refT100}
{\author{M.C.}{Torrent} \and \author{M.}{San Miguel}}{1988}
           {Stochastic-dynamics characterization of delayed laser
threshold instability with swept control parameter}
           {Phys. Rev. A}{38}{245--251}

\refbook{refB200}
{\author{E.}{Beno\^\i t (Ed.)}}{1991}
            {Dynamic bifurcations}
            {Springer}{Berlin}

\refbook{refG100}
{\author{C.W.}{Gardiner}}{1990}
    {Handbook of Stochastic Methods}
    {Springer}{Berlin}
\refarticle{refB100}
{\author{G.}{Broggi}, \author{A.}{Colombo}, \author{L.A.}{Lugiato}
\and \author{Paul}{Mandel}}{1986}
          {Influence of white noise on delayed bifurcations}
           {Phys. Rev. A}{33}{3635--3637}

\endgroup
\inrefsfalse

\spaceandathird

\damtpaddress


\windowcover

\def\theauthor{G.D.~Lythe and M.R.E.~Proctor}

\title{ Noise and slow-fast dynamics in  a three-wave resonance
problem}

\signature

\abstract{
Recent research on the dynamics of certain fluid
dynamical instabilities shows that when there is a slow invariant manifold
subject
to fast timescale instability the dynamics are extremely sensitive to 
noise.
 The behaviour of such systems can be described in terms of a
one-dimensional map, and previous work has shown how the effect of noise can
be modelled by a simple adjustment to the map. Here we undertake an in depth
investigation of a particular set of equations, using the methods of
stochastic
integration. We confirm the prediction of the earlier studies that the noise
becomes important when $\mu |\ln\epsilon| = \Or{1}$, where $\mu$ is the
small timescale ratio and $\epsilon$ is the noise level. In addition, we
present detailed information about the statistics of the solution when the
noise is a dominant effect; the analytical results show excellent agreement
with numerical simulations.
}

\publishedin{Phys.Rev.E {\bf 47} 3122-3127 (1993)}

\begingroup
\ShowTP
\endgroup

\pageno=1

\beginsect Introduction

In many circumstances, a low order system of
 ordinary differential equations (ODEs)
serves as a useful model for a physical system.  A difficulty
arises, however, if the solutions are noticeably affected by
small external noise.  This is the case in
several systems of physical interest sharing the characteristic that their
solutions  consist of alternating slow and fast 
phases\addthreerefs{refH101}{H101}{refH102}{H102}{refH103}{H103}.
In this paper we take as our example the following third order system
of ODEs describing the resonant interaction of three wave
modes when one mode is unstable and the other two 
damped\addthreerefs{refV100}{V100}{refW100}{W100}{refB102}{B102}:
$$\eqalign{\dot x &= \mu x -  y^2 +2z^2 - \delta z,\cr
           \dot y &= y(x-1),\cr
           \dot z &= \mu z + \delta x - 2xz. }
\eqno \myeqno \myeqlabel{ftwc}$$

When $\mu$
 is small the character of solutions of~\ftwc\ is dramatically changed
by tiny amounts of additive noise -- the bifurcation structure
 with a full gamut of periodic orbits
and chaotic regions is replaced by a noisily periodic orbit across a
wide range of parameter values (Figure 1).
The quantity $\mu$ is the ratio of the
instability of the unstable mode (its exponential rate of growth in
the absence of interaction) to the damping rates of the other two
modes (assumed equal). \addref{refNote1}{Note1}
\figure 8truein 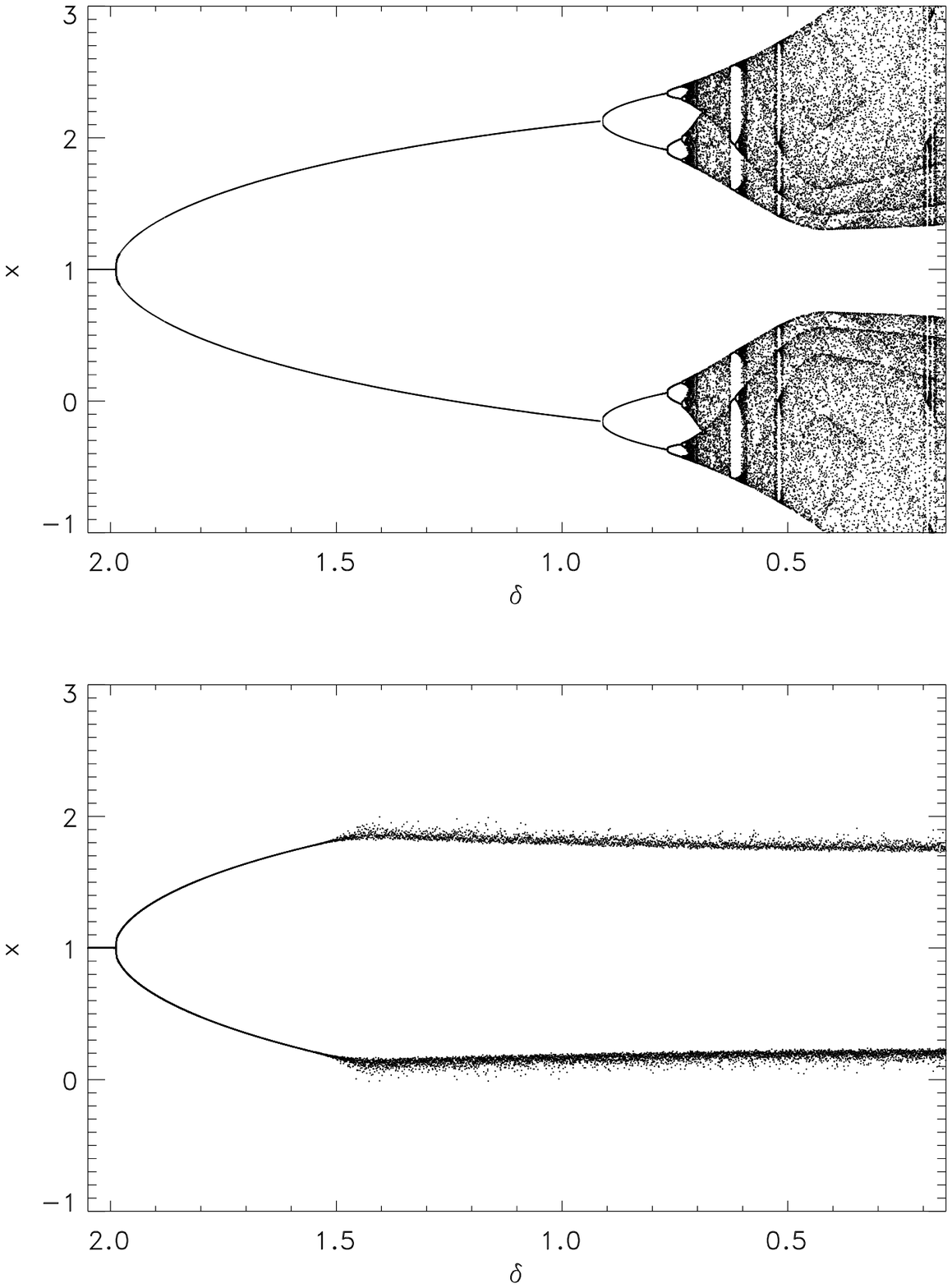
{\it Bifurcation diagrams with and without noise for $\mu=0.01$.}
The graphs are obtained from numerical simulation of~\ftwc\ and
each dot represents a turning point of $x$.  The top graph is obtained
with no added noise; the bottom graph with very small
noise (r.m.s magnitude $\epsilon=10^{-10}$) added to the variable $y$.
The noiseless bifurcation structure including chaotic regions is
replaced by a noisily periodic orbit for $\delta < 1.5$.
\myfigurelabel{bif}

\figure 5truein  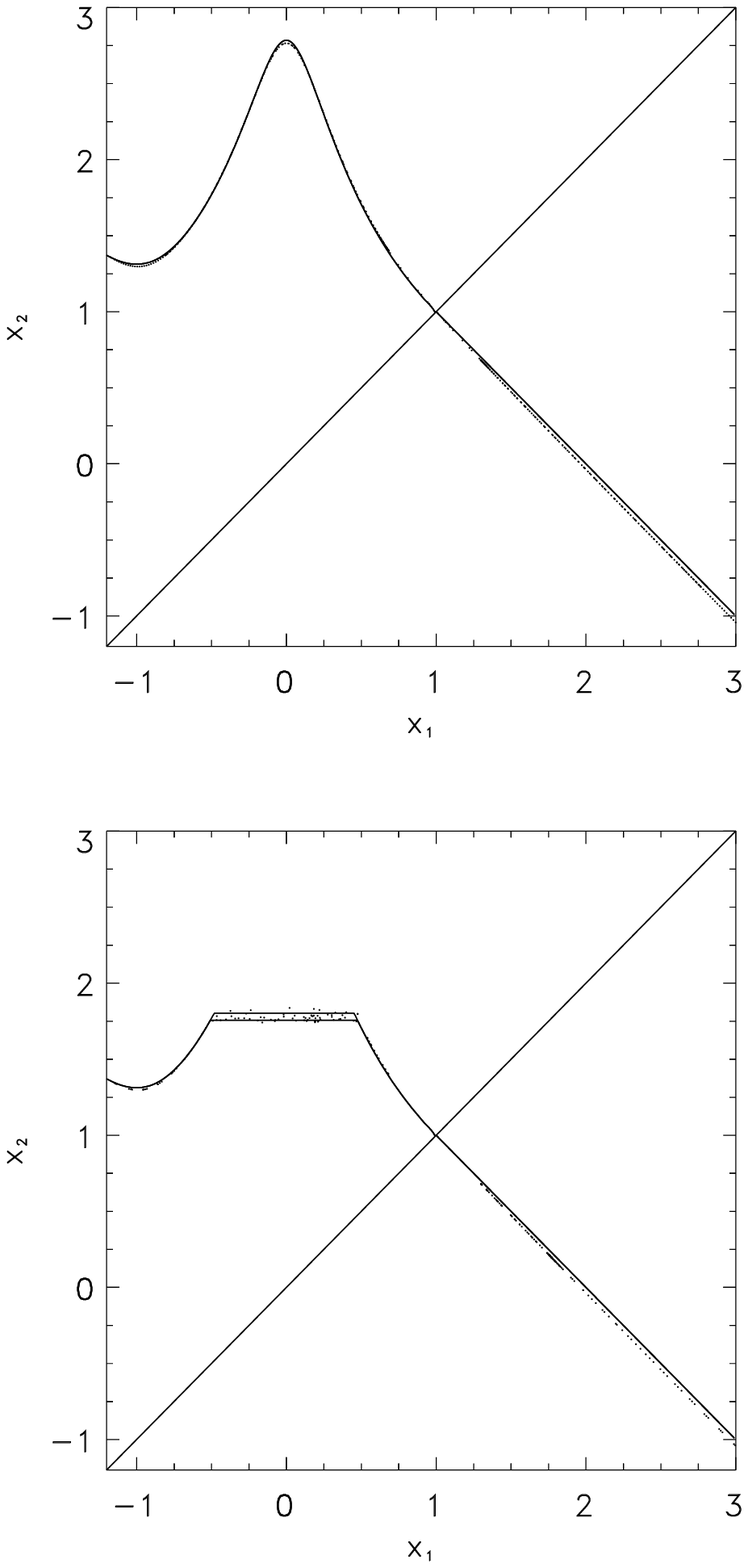
{\it The one-dimensional map.}
\mapexpl  In the top figure, the solid line is the formula of Hughes and
Proctor\addref{refH101}{H101} for $\mu=0.01$ and $\delta=0.5$. The dots
are obtained from numerical solution of~\ftwc .  The same map is
shown below with a flat top due to noise. 
 The two lines in the flat top are $\mean{\xm} + \sigma_{\xm}$
and $\mean{\xm} - \sigma_{\xm}$, calculated as described in the text with
$\epsilon=10^{-10}$, and the dots are numerical results with white noise of
magnitude $10^{-10}$ added to the variable $y$.
\myfigurelabel{figmap}

It is possible to describe the dynamics of~\ftwc\ in terms of
a one-dimensional map. As shown in\addref{refH101}{H101},
analytical expressions can be obtained for this
map by assuming
that the solutions consist of alternating slow and fast phases, and
solving approximations to~\ftwc\ in each phase (\figmap ).
In the slow phase the system is close to 
the invariant plane $y=0$ and
moves slowly from the region
where the plane is attracting ($x<1$) to the region where it is
repelling ($x>1$).  This phase is occasionally interrupted by a fast phase
which reinjects the system close to the attracting part of $y=0$.
  It is in the slow phase that the sensitivity to noise arises.

Here we use a stochastic differential equation 
to calculate the adjustment to the map of Hughes and
Proctor\addref{refH101}{H101} necessary 
to describe the dynamics of the slow phase in the
presence of additive white noise.
  The smallness of $\mu$ is
responsible for the division of the dynamics into two phases and the
sensitivity to noise.  In this paper, all calculations are done to
lowest order in $\mu$.

The map we use is a map of successive turning points of
$x$.  In the presence of noise the turning point of $x$ which defines
the end of the slow phase is a random variable, $\xm$.  In the heuristic model of Hughes and Proctor\addref{refH101}{H101},
the noise determines $\xm$ if $\mu|\ln\epsilon | < \Or{1}$,
where $\epsilon$ is the r.m.s. noise level, and $\xm$ is
estimated by assuming that $|y|=\Or{\epsilon}$ at $x=1$.  
The value of $\xm$ calculated in this manner is $\Or{1}$ less than the
corresponding deterministic value.

In this paper, we extend the treatment of Hughes and Proctor using a
stochastic differential equation to describe the slow phase.  We
derive explicit expressions for the \pd\ of $\xm$, and for the
condition on $\mu|\ln\epsilon |$ which marks the transition to the
noise-controlled \regime .  In this \regime , 
the most probable value of $\xm$ is a function of $\mle$ and the
standard deviation of its \pd\ is proportional to $\mu$.
For small $\mu$, our calculations predict accurately the results of
numerical solution of~\ftwc\ with low-level noise added to the variable $y$.
The numerical results presented here were obtained using a simple
extension of the Heun (second order Runge-Kutta)
 method for integrating ODEs to include additive
white noise\addtworefs{refH104}{H104}{refK100}{K100}.
  The increment to $y$ at
each step includes a Gaussian random variable proportional to
$\epsilon$ and to the square root of the timestep.

Slow-fast dynamics similar to those of the three-wave resonance system are 
 relevant in other contexts.  Our results are presented in a such a way
that they can be easily generalised.  In the appendix, we show how a
generic slow phase is reduced to a problem taken from dynamic bifurcation
theory, and summarise some results for this case.

\beginsect The slow phase with and without white noise

The slow phase of~\ftwc\ is defined as beginning when the following are true:
$$y^2 \ll \mu ,\qquad
            z={\delta \over 2} + \Or{\mu},\qquad{\rm and}\qquad
            \Or{\mu} < x < 1.\eqno \myeqno \myeqlabel{slowstart}$$
We then find that~\ftwc\ reduce to:
$$\eqalign{\dot x &= \mu f(x)-y^2,\cr
\dot y &= yg(x),\cr
z &\simeq {\delta \over 2} + \mu{\delta \over 4x}}
\eqno \myeqno \myeqlabel{genslow}$$
$${\rm where}\quad f(x)=x+{\delta^2 \over 4x}\quad {\rm and}\quad g(x)=x-1.
\eqno \myeqno \myeqlabel{fxgx}$$
When the initial conditions $(x_0,y_0,z_0)$ satisfy~\slowstart\ we
observe the following:\break
\noindent  -the variable $y$ decreases exponentially until
$x=1$ and then increases exponentially;\break
\noindent -the variable $x$ is the
driving variable, evolving independently until the
very end of the slow phase;\hfil\break
\noindent -the remaining variable, $z$, is of
secondary importance for $x>\Or{\mu}$ because it is `slaved' to $x$ (given as a
function of $x$).

Our strategy for determining $\xm$ is to take $\dot x =
\mu f(x)$, so that $x$ is a function of time, and then solve for $y$
as a function of time. $\xm$ is then the value
of $x$ at which $y^2 = \mu f(x)$.  The term $-y^2$ in the equation for
$\dot x$ has a small effect at the end of the slow phase
which we calculate a correction for.

In the presence of white noise $y$ becomes a stochastic process $y_t$ satisfying
the following stochastic differential equation:
$$dy_t = y_t\tilde g(t)dt + \epsilon dW_t \eqno \myeqno \myeqlabel{genysde}$$
where $\tilde g(t) = x(t) -1$, $\epsilon$ is a constant with
$0\le\epsilon \ll \mu$ and $W_t$ is the Wiener process.
\nom{$W_t$}{Wiener Process}
  Exact solution of~\genysde\ is 
possible\addref{refG100}{G100}:

$$y_t = G(t,t_0)\left (y_0 + \epsilon\int_{t_0}^t{1 \over
G(s,t_0)}dW_s\right ) \eqno \myeqno \myeqlabel{gensdesoln}$$
where
$$G(t,t_0) = e^{\int_{t_0}^t \tilde g(u)du}. \eqno \myeqno
\myeqlabel{biggdef}$$ 
The mean value of $y_t$,
$$\bigl<y_t\bigl> = G(t,t_0)y_0, \eqno \myeqno \myeqlabel{genmeany}$$
is the solution in the limit $\epsilon \to 0$.
The \pd\ of $y$ is Gaussian with standard deviation, $\sigma_y$, a
function of time given by
$$\sigma_y^2 = \bigl<y_t^2\bigl> - \bigl<y_t\bigl>^2 = \epsilon^2 G^2(t,t_0)\int_{t_0}^t{1 \over G^2(s,t_0)}ds .\eqno \myeqno \myeqlabel{genmeanys}$$
If $x=1$ at $t=t_{\alpha}$ then for $t-t_{\alpha} > \Or{{1 \over \sqrt{\mu}}}$
$$\sigma_y^2 \simeq \epsilon^2 \sqrt{\pi a}G^2(t,t_{\alpha})
\eqno \myeqno \myeqlabel{yes}$$
where
$$a={1 \over \mu f(1)g'(1)}.
\eqno \myeqno \myeqlabel{adef}$$
\beginsubsect{The deterministic limit}

If $\sigma_y \ll \left< y_t \right>$ for $x>1$ then the noise can be
treated as 
a small perturbation to the deterministic solution given
by~\genmeany\ or by 
$$y = y_0 e^{-{1 \over \mu}(F(x)-F(x_0))}
 \eqno \myeqno \myeqlabel{sitox}$$
where
$$ F(x)={1 \over 2}\ln (1+{4x^2 \over \delta^2}) -x +
{\delta \over 2}\tan^{-1}\left({2x \over \delta}\right).
 \eqno \myeqno \myeqlabel{fxtwc}$$
The \pd\ of $\xm$ in this case is very narrow and the
mean value, $\mean{\xm}$, satisfies
$$F(\left< \xm \right>) - F(x_0) = \mu\left(\ln y_0 - {1 \over 2}\ln(\mu f(\mean{\xm})\right). \eqno \myeqno \myeqlabel{detxmax}$$
\beginsubsect{The noise-controlled \regime\ }

If $\sigma_y \gg \left<y_t\right>$ for $x>1$ then it is the noise
rather than the initial conditions
which controls $\xm$.  In this case
the \pd\ of $y$ for $x>1$ is Gaussian with negligible mean and
exponentially rising standard deviation.
Equation~\yes\ corresponds to the fact that,
for $t-t_{\alpha} > \Or{{1 \over \sqrt\mu}}$,
 the path of any one realisation is
approximately deterministic (but starting from a level which is random variable).

Knowing $\sigma_y$ as a function of time, we can calculate the
probability that  $y^2$ is greater than $\mu f(x)$ at any time.  The \pd\ of $\xm$
is the derivative with respect to $x$ of this probability.
The probability that $\xm$ lies between $x$ and $x+dx$ is therefore
$R(x)dx$ where
$$\eqalign{R(x) &= 2{\partial \over \partial x}{1 \over\sqrt{2\pi}\sigma_y}
\int_{\scriptscriptstyle-\infty}^{\scriptscriptstyle-\sqrt{\mu f(x)}}
\displaystyle e^{-{y^2 \over 2\sigma_y^2}}dy\cr
&\simeq{2 \over \sqrt{2\pi}}{g(x) \over \mu f(x)}
{\sqrt{\mu f(x)} \over \sigma_y}e^{-{\mu f(x) \over 2 \sigma_y^2}}.}
\eqno \myeqno \myeqlabel{genrx}$$
The maximum value of $R(x)$ is at $x=\hat x$ where $\hat x$ is defined
by the condition $\sqrt{\mu
f(\hat x)}=\sigma_y$. (Both $x$ and $\sigma_y$ are functions of time.)
Thus $\hat x$ satisfies
$${\sqrt{\mu f(\hat x)} \over  \exp\left({\int_{1}^{\hat x}{g(u)
\over \mu f(u)}du}\right)}=\epsilon(\pi a)^{1 \over 4}
\eqno \myeqno \myeqlabel{xhatsat}$$
or equivalently
$$F(1)-F(\hat x) = \mu |\ln\epsilon | + {\mu \over 2}\ln (\mu f(\hat x))
-{\mu \over 4}\ln(\pi a) .
\eqno \myeqno \myeqlabel{nicerxhatsat}$$
At the very end of the slow phase the simple relationship $\dot x=\mu f(x)$
 breaks down because $y^2$ is no longer negligible.
We obtain a more accurate expression for the most probable value of
$\xm$ by replacing $\hat x$ by
$\xhc = \hat x - \Delta \hat x$ where
$$\Delta \hat x = \int^{\xm} y^2 {dt \over dx}dx \quad
        \simeq \quad {1 \over 2}{\mu f(\hat x) \over g(\hat x)}.
\eqno \myeqno \myeqlabel{genxhatcorr}$$

We exhibit the \pd\ of $\xm$ by defining the random variable $v$ as
$$v=-(\xm-\xhc){g(\xhc) \over \mu f(\xhc)}.
\eqno \myeqno \myeqlabel{vdefn}$$
The probability that $v$ lies between $v$
and $v+dv$ is $\tilde R(v)dv$ where
$$\tilde R(v)\simeq {2 \over \sqrt{2\pi}}\,e^v\,e^{-{1 \over 2}e^{2v}}
\eqno \myeqno \myeqlabel{rtilde}$$
which is the \pd\ of the log of the absolute value of a Gaussian
random variable with unit variance.
  Explicit expressions for the mean and variance of $v$ exist:
$$\bigl<v\bigl>=-{1 \over 2}(\gamma + \ln 2)
\eqno \myeqno \myeqlabel{meanv}$$
where $\gamma=.577 \ldots$ (Euler's constant) and
$$ \bigl<v^2\bigl>-\bigl<v\bigl>^2 = {\pi^2 \over 8}\eqno \myeqno \myeqlabel{meanvs}$$
so the mean and \sd\ of $\xm$ in the noise-controlled \regime\ are given by
$$\mean{\xm} = \xhc +{1 \over 2}{\mu f(\xhc) \over g(\xhc)}(\gamma +
\ln 2)
\eqno \myeqno \myeqlabel{meanxmax}$$
$$ \sigma_{\xm}=\sqrt{\bigl<\xm^2\bigl>-\bigl<\xm\bigl>^2}={\pi \over
2\sqrt{2}}{\mu f(\xhc) \over g(\xhc)} .
\eqno \myeqno \myeqlabel{sdxm}$$
Each value of $x_1$ for $-1+\Or{\sqrt{\mu}} < x_1 < 1$ in~\figmap\ corresponds to
a set of initial conditions for the slow phase\addref{refH101}{H101}. 
 The value taken for  $x_2$ is the lower of the two values given by
 ~\detxmax\ and ~\meanxmax . When the noise-controlled
value~\meanxmax\ is the lower, we plot $\mean{\xm}+\sigma_{\xm}$
and $\mean{\xm}-\sigma_{\xm}$.  For the purposes of~\figmap , 
the transition from the
noise-controlled to the deterministic \regime\ is sufficiently rapid
that it is unnecessary to consider the transition region.
  In the next section, however, we derive a more
general formula for the \pd\ of $\xm$.
\figure 5truein  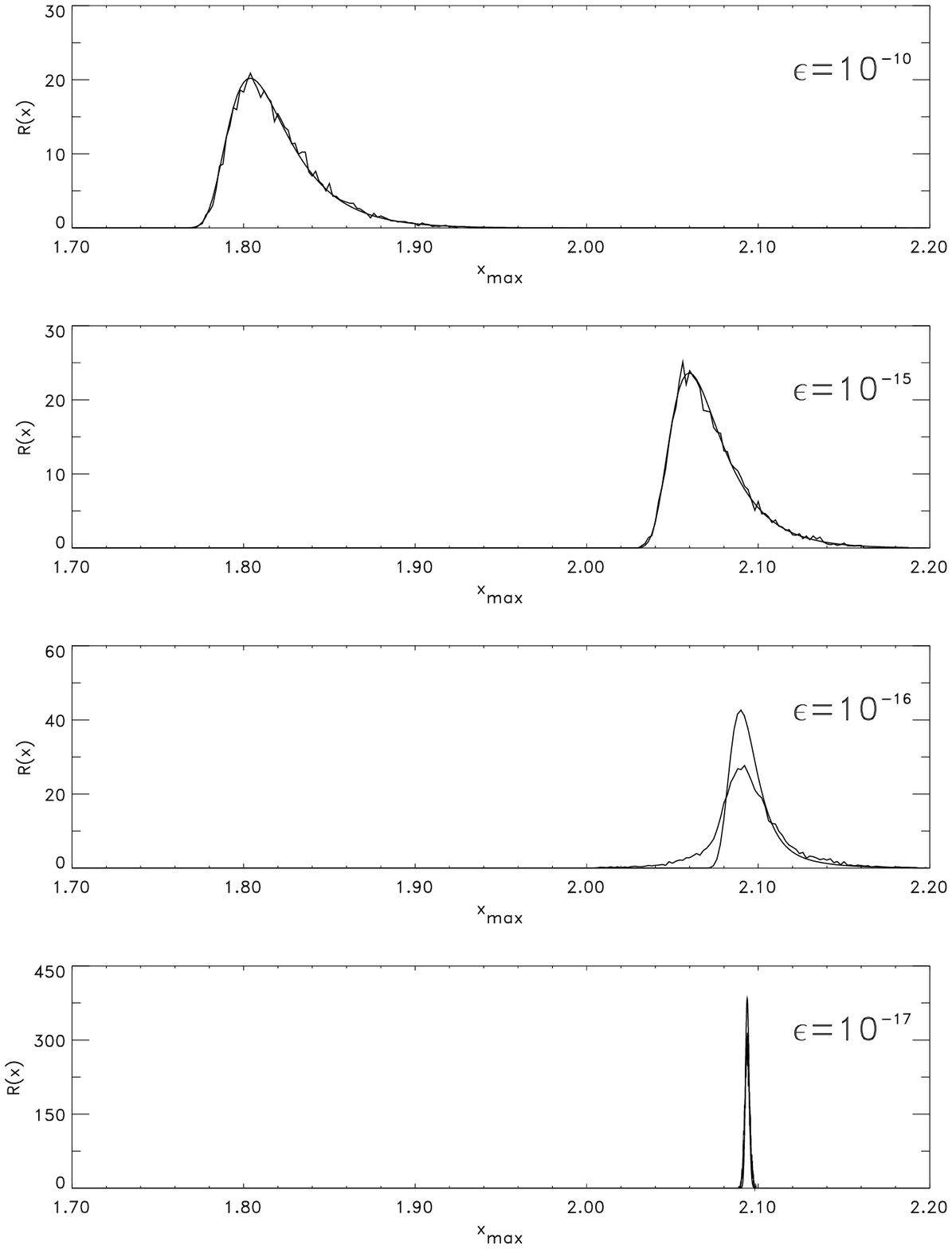
{\it The effect of noise on the probability distribution of} $\xm$.
Numerically-obtained \pd s of $\xm$ with $\mu=0.01$ and
$\delta=1.0$.
The smooth curve in each case is the function $R(x)$ (27) which
reduces to~\genrx\ in the noise-controlled \regime .
\myfigurelabel{figshape}

\figure 5truein 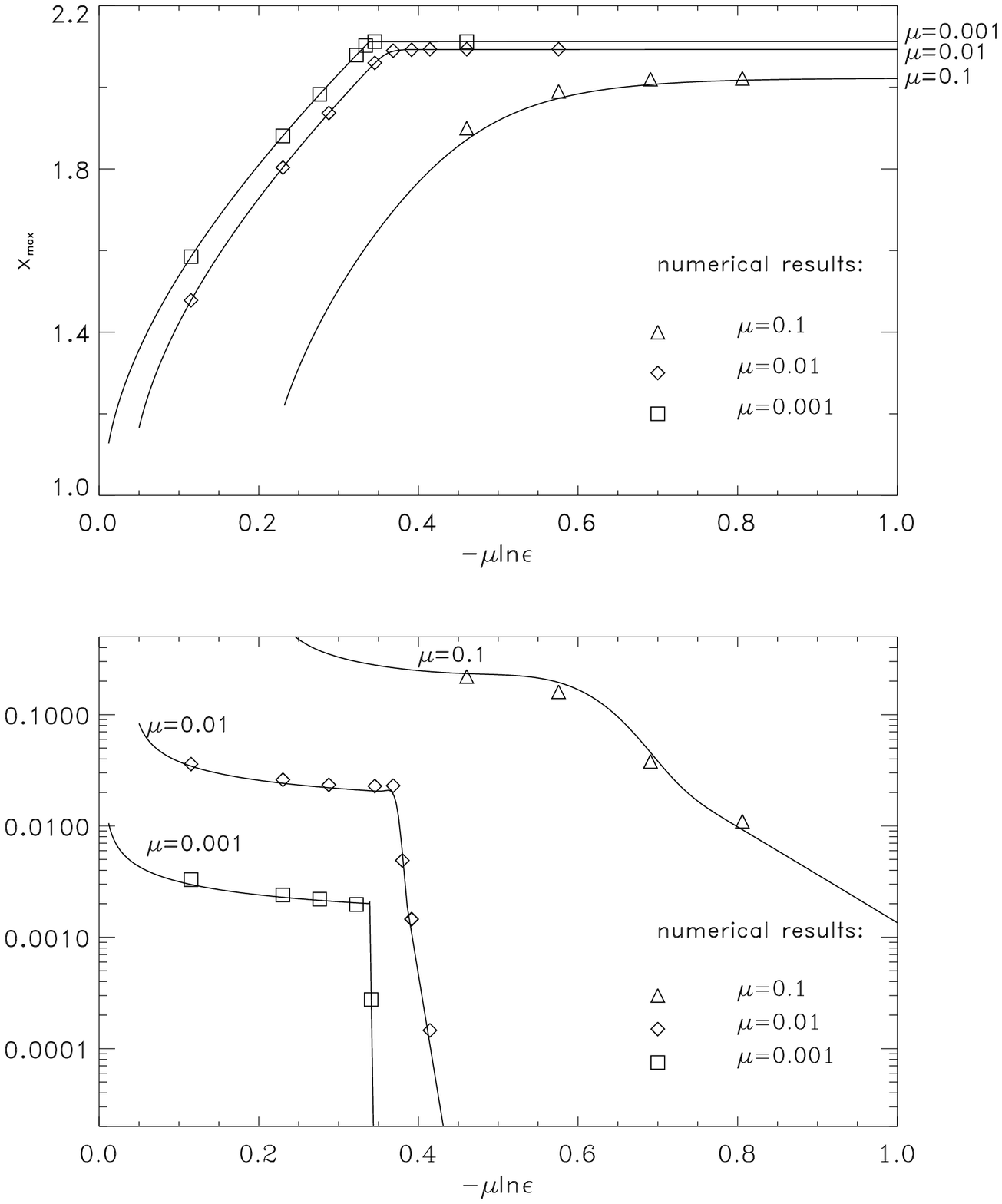
{\it Mean and standard deviation of $\xm$ as
 a function of $\mle$.}
The curves are predictions obtained from the explicit form for the
\pd\ of $\xm$ (27) and some numerical results are shown.
The agreement between small-$\mu$ calculations and numerical results
is good even for $\mu$ as large as 0.1. The knee in the graph of
$\sigma_{\xm}$ vs $\mle$ corresponds to the end of the
noise-controlled \regime .
\myfigurelabel{pred}

\beginsubsect{A more general formula}

We express the relative magnitudes of the 
deterministic and noise-driven parts of $y_t$ via the parameter $c$
defined as
$$c={\mean{|y_t|} \over \sigma_y}
\eqno \myeqno \myeqlabel{cdef}$$
which is constant for $t-t_{\alpha} > \Or{{1 \over \sqrt{\mu}}}$~\yes .
The noise-controlled \regime\ corresponds to $0<c\ll 1$, and the
deterministic limit to $c \gg 1$.  The condition $c<1$ for $t-t_{\alpha}
> \Or{{1 \over \mu}}$ can be taken as a
test of whether the noise-controlled \regime\ is in force.  This
condition is
$$\mu |\ln \epsilon | < F(1)-F(x_0)+
\mu \left({1 \over 4}\ln(\pi a)-\ln y_0 \right ).
\eqno \myeqno \myeqlabel{mlecond}$$

A more general formula for $R(x)$ is obtained by allowing the \pd\ of
$y$ for $x>1$ to have non-zero mean.  Thus
$$\eqalign{R(x) &= {\partial \over \partial x}{1 \over
\sqrt{2\pi}\sigma_y}\int_{\scriptscriptstyle-\sqrt{\mu f(x)}}^{\scriptscriptstyle\sqrt{\mu f(x)}}
e^{-{(y-\mean{y})^2 \over 2\sigma_y^2}}dy\cr
&\simeq{1 \over \sqrt{2\pi}}\,u\left(e^{-{1 \over 2}(u-c)^2} + e^{-{1
\over 2}(u+c)^2}\right){g(x) \over \mu f(x)}}
\eqno \myeqno \myeqlabel{moregenrx}$$
where
$$\qquad u={\sqrt{\mu f(x)} \over \sigma_y}.
 \eqno \myeqno \myeqlabel{ucdef}$$
This \pd\ is compared with numerical results for $\mu=0.01$
in~\figshape . In the limit $c\to 0$ we find the 
 distribution \rtilde.
For large $c$ we regain the deterministic \regime\ (narrow,
Gaussian distribution of $\xm$).

Our results are exact for small $\mu$ and delta function initial
conditions.  To produce~\figshape\ and~\pred\ we take the initial
conditions from the corresponding deterministic orbit.
This gives excellent agreement in the noise-controlled \regime , where
$\xm$ is independent of initial conditions, and gives the correct
large-$c$ limit for $\mean{\xm}$. However the standard deviation
$\sigma_{\xm}$ is underestimated
  for nonzero $c$ because the \pd\ of $\xm$ is carried through the fast
phase, so the initial conditions for the slow phase vary from cycle to
cycle in the noisily periodic orbit.  The broadening of the \pd\ of
$\xm$ produced by this is most noticeable in the transition region
between the noise-controlled and deterministic \regime s
($\epsilon=10^{-16}$ in~\figshape).

In~\pred\ we compare numerical results for the most probable value of
$\xm$ and the standard deviation of $\xm$ with values calculated
using the explicit form of the \pd\ ~\moregenrx .  The most probable value of $\xm$ we find from
 the (approximate) condition
$$u^2 = 1 + c^2.
\eqno \myeqno \myeqlabel{useopcs}$$
This corresponds to $\mean{y_t^2} =\mu f(x)$
and gives the correct result in the large and small $c$ limits.
The standard deviation of $\xm $ can be written
$$\sigma_{\xm} \simeq {\pi \over 2\sqrt{2}}
{\mu f(x) \over g(x)}h(c)
\eqno \myeqno \myeqlabel{moregensd}$$
where $f(x)$ and $g(x)$ are evaluated at the most probable value of
$\xm$, $h(0)=1$ and, for large $c$, $h(c)\simeq{1 \over c}$.
The form we have used for $h(c)$ in~\pred\ is $h^2(c)={1 \over 1+c^2}+
c^2e^{{-c^2 \over 1.5}}$, which we obtained as a fit to numerical
integration of~\moregenrx .

\beginsect{Conclusion}

Slow-fast dynamical systems such as the three-wave resonance system
discussed here are most conveniently described in terms of a
one-dimensional map.  Low-level white noise has an $\Or{1}$ effecttrue
which can be calculated by solving a stochastic differential equation.
In the deterministic limit $\xm$,
the turning point of $x$ which defines
the end of the slow phase, is determined by the initial conditions.
In the noise-controlled \regime , which is in force when $\mu
|\ln\epsilon |$ is less than an $\Or{1}$ constant which depends on
initial conditions, $\xm$ is a random variable with 
most probable value a function of $\mle$ and
standard deviation proportional to $\mu$.

We know of several other physical contexts which give rise to
noise-sensitive slow-fast dynamics.  One is the shear instability of
tall thin convection cells\addtworefs{refH102}{H102}{refH103}{H103}.
Another is pulsating laser oscillations%
\addtworefs{refM100}{M100}{refM104}{M104}
consisting of short pulses separated by long periods of very small
intensity. A related problem is
that of random perturbations of heteroclinic
attractors\addref{refS100}{S100}, where noise controls the
length of time spent near an unstable fixed point, and a stable
homoclinic or heteroclinic orbit
provides the reinjection.

\beginack
\noindent G.L. is grateful for financial support from the
Commonwealth Scholarship Commission.

\beginappendix Appendix

In the three-wave resonance problem considered above, the slow phase
is defined by
$$\eqalign{\dot x &= \mu f(x),\cr
dy_t &= g(x)y_tdt + \epsilon dW_t}
\eqno \myeqno \myeqlabel{genslow}$$ 
where $f(x)=x+{\delta^2 \over 4x}$ and $g(x)=x-1$.  We obtain a
 system often studied in dynamic bifurcation theory\addref{refB200}{B200}
if we put $f(x)=1$ and $g(x)=x$.  Then $y_t$ satisfies
$$dy_t = g(t)y_tdt + \epsilon dW_t \qquad \hbox{\rm where}\quad g=\mu
t
\eqno \myeqno \myeqlabel{db}$$ 
Suppose $y=y_0$ for some $t=t_0<0$.  (In the
three-wave resonance problem the natural choice is 
$y_0=\Or{\sqrt{\mu}}$ and $-\mu t_0=\Or{1}$.)
The next value of $g$ at
which $y=y_0$, $\bar g$,  defines a dynamic bifurcation point, occurring later than
the `static' bifurcation at $g=0$.  In fact, in the deterministic limit,
$\bar g = -g(t_0)$.  

The noise-controlled \regime\ is in force if
$$\mu|\ln \epsilon| <
{1 \over 2}g^2(t_0) +\mu \left ( {1 \over 4}\ln{\pi \over \mu}-\ln y_0
\right ).
\eqno \myeqno \myeqlabel{dbtrans}$$ 
In this \regime , $\bar g$ is a random variable with
most probable value, $\hat g$, given by 
$$\hat g^2 = 2\mu |\ln \epsilon| +{\mu \over 2}\ln{\mu \over \pi}+2\mu
\ln y_0
\eqno \myeqno \myeqlabel{hgs}$$ 
and standard deviation $\sigma_{\bar g}$ by
$$\sigma_{\bar g} = \mu {\pi \over 2\sqrt{2}}{1 \over \hat g}.
\eqno \myeqno \myeqlabel{sbg}$$ 
In this form, our results appear consistent with results obtained
numerically\addref{refB100}{B100}, from an electronic circuit model of a
ring laser\addref{refM102}{M102}
and with analytical results for the laser 
threshold instability\addref{refT100}{T100}.

\showrefs
\end